\documentclass[a4paper,11pt]{article}

\begin{document}

\title{A class of colliding waves in metric--affine gravity,
nonmetricity and torsion shock waves}

\author{
Alfredo Mac\'{\i}as$^{\star \diamond}$\thanks{E-mail: macias@fis.cinvestav.mx,
amac@xanum.uam.mx},
Claus L\"ammerzahl$^{\P}$\thanks{E-mail: claus@spock.physik.uni-konstanz.de},
and Alberto Garc\'{\i}a$^\star$\thanks{E-mail: aagarcia@fis.cinvestav.mx}\\
$^{\star}$  Departamento de F\'{\i}sica, CINVESTAV--IPN,\\
Apartado Postal 14--740, C.P. 07000, M\'exico, D.F., MEXICO\\
$^{\diamond}$ Departamento de F\'{\i}sica,
Universidad Aut\'onoma Metropolitana--Iztapalapa,\\
Apartado Postal 55-534, C.P. 09340, M\'exico, D.F., MEXICO.\\
$^{\P}$Fakult\"at f\"ur Physik, Universit\"at Konstanz \\
Postfach 5560 M674, D--78434 Konstanz, Germany}

\date{\today}

\maketitle

\begin{abstract}
By using our recent generalization of the colliding waves
concept to metric--affine gravity theories, and also our generalization of the 
advanced and retarded time coordinate representation in terms of Jacobi functions, 
we find a general class of colliding wave solutions with fourth degree  
polynomials in
metric--affine gravity. We show that our general approach contains the
standard second degree polynomials colliding wave solutions as a particular
case. {\bf file shock7b.tex 19.05.2000}

\end{abstract}
\vspace{0.5cm}


\section{Introduction}

Recently we generalized \cite{glmms98} the formulation of the general
relativistic colliding gravitational waves concept
\cite{khan,szek,chandra1,chandra2} to metric--affine gravity theories (MAG)
and presented a simple example of such kind of exact solutions.
The resulting plane waves are equipped
with five symmetries and the resulting geometry after the collision possesses 
at least two spacelike Killing vectors.
The solution presented describes the scattering of two noncollinear polarized 
gravitational plane waves. At the leading edge of each colliding type--N
gravitational wave, the curvature tensor exhibits a jump discontinuity
arising, for example, from the second derivative
$(-U^2)^{\prime\prime} = u^2 \delta^\prime(u) + 4 u \delta(u) + 2 \Theta(u)$.  
The former is interpreted as a gravitational impulsive
wave, whereas the latter is attributed to a gravitational shock wave.

As far as the nonmetricity and torsion are concerned, they could present  
delta singularities and
jump discontinuities. However, even then the Bianchi identities hold in a
distributional sense \cite{taub}.
In particular, also $D T^\alpha = R_\beta^{\phantom{\alpha}\alpha}\wedge
\vartheta^\beta$ holds.
There are no problems on the right--hand side because the delta type
singularities of the curvature are multiplied by the smooth distributions
$\sqrt{1 - \Theta(u) u^2}$ and $\sqrt{1 - \Theta(v) v^2}$, respectively.
Thus, one can interpret them as torsion and nonmetricity gravitational
shock waves.

All techniques for generating solutions can be applied to generate
cylindrically symmetric spacetimes, in particular colliding plane waves.
The way of derivation of these solutions is related to the search of a class 
of cylindrically symmetric solutions in MAG, starting with the line element
\cite{honnef}
\begin{equation}
ds^2= \frac{1}{H(x,t)}\left[\widetilde{\Delta} \left( \frac{dx^2}{X(x)}
- \frac{dt^2}{\Upsilon(t)}\right) +
\frac{X(x)}{\widetilde{\Delta}}\left(dy + \widetilde N(t)d\sigma\right)^2
+ \frac{\Upsilon(t)}{\widetilde{\Delta}}\left(dy + \widetilde M(x)d\sigma
\right)^2\right]\, ,
\end{equation}
with $\widetilde{\Delta}:= \widetilde M - \widetilde N$.
It is clear that $t$ is a timelike coordinate, while $x$, $y$, and $\sigma$
are spacelike ones. Notice that $\partial_y$ and $\partial_\sigma$ are
spacelike Killing vectors. Moreover, since we are working with signature
$(+,+,+,-)$ the ranges of the $x$-- and $t$--variables require the fulfillment
of the conditions $X(x)>0$, $\Upsilon(t)>0$, $\widetilde{\Delta}>0$, and  
$H(x,t)>0$ in
order to mantain the signature. Therefore, this signature condition
yields constraints on the coefficients appearing in the polynomials.

Assuming first that
$X(x)$ and $\Upsilon(t)$ are polynomials up to fourth degree
on $x$ and $t$, respectively, second $\widetilde M$ and $\widetilde N$ are
polynomials up to second degree also on $x$ and $t$, respectively, and third the 
torsion and nonmetricity are proportional rational functions. Therefore,
one arrives at algebraic equations, solvable by computer algebra programs,
for the polynomials coefficients. Moreover, it is always possible to
introduce the retarded and
advanced time coordinates $u$ and $v$. However, only certain solutions
satisfy the Ernst \cite{ernst} requirement for being colliding waves.

The main  pruporse of this paper is to give the formulation in MAG of the
most general class of colliding waves, nowadays known, i.e., with fourth
degree polynomials .
As usual, it is assumed that in the corresponding spacetime, the two
waves approach each other, from opposite sides, in flat Minkowski background.
After the collision, a new gravitational field equipped with torsion and
nonmetricity evolves, which satisfies certain continuity conditions. The
colliding plane waves possess five symmetries, while the geometry resulting
after the collision has two spacelike Killing vectors.

The plan of the paper is as follows: In Sec. II the general quadratic MAG
Lagrangian is revisited. In Sec. III a class of cylindrically symmetric waves
in MAG is presented. In Sec. IV we review briefly the generalization of
the colliding waves concept to MAG and the general representation
through advanced and retarded time coordinates for fourth degree polynomials 
is introduced. In Sec. V we reduce the general approach to
second degree polynomials and the results are discussed. In Sec. VI we
outlook the theory.

\section{General quadratic MAG Lagrangian}

In a metric--affine spacetime \cite{PR}, a general parity conserving
quadratic Lagrangian reads:
\begin{eqnarray}
\label{QMA} V_{\rm MAG}&=&
\frac{1}{2\kappa}\,\left[-a_0\,R^{\alpha\beta}\wedge\eta_{\alpha\beta}
-2\lambda\,\eta+T^\alpha\wedge{}^*\!\left(\sum_{I=1}^{3}a_{I}\,^{(I)}
T_\alpha\right)\right.\nonumber\\ &+&\left.
2\left(\sum_{I=2}^{4}c_{I}\,^{(I)}Q_{\alpha\beta}\right)
\wedge\vartheta^\alpha\wedge{}^*\!\, T^\beta + Q_{\alpha\beta}
\wedge{}^*\!\left(\sum_{I=1}^{4}b_{I}\,^{(I)}Q^{\alpha\beta}\right)\right.
\nonumber \\&+&
b_5\bigg.\left(^{(3)}Q_{\alpha\gamma}\wedge\vartheta^\alpha\right)\wedge
{}^*\!\left(^{(4)}Q^{\beta\gamma}\wedge\vartheta_\beta \right)\bigg]
\nonumber\\&- &\frac{1}{2}\,R^{\alpha\beta} \wedge{}^*\!
\left(\sum_{I=1}^{6}w_{I}\,^{(I)}W_{\alpha\beta} +
\sum_{I=1}^{5}{z}_{I}\,^{(I)}Z_{\alpha\beta}\right)
\label{lobo}\,.
\end{eqnarray}
The signature of spacetime is $(-+++)$, $\eta:={}^*\!\, 1$ is the volume
four--form and the constants
$a_0,\cdots a_3$, $b_1,\cdots b_5$, $c_2, c_3,c_4$,
$w_1,\cdots w_6$, $z_1,\cdots z_5$ are dimensionless.
We have introduced in the curvature square term
the irreducible pieces of the antisymmetric part
$W_{\alpha\beta}:= R_{[\alpha\beta]}$ and the symmetric part
$Z_{\alpha\beta}:= R_{(\alpha\beta)}$ of the curvature two--form.
It is worthwhile to note that in $Z_{\alpha\beta}$, we meet a purely
post--Riemannian part.

The first and the second field equations of MAG \cite{PR} read
\begin{eqnarray}
DH_{\alpha}- E_{\alpha}&=&\Sigma_{\alpha}\,,\label{first}\\
DH^{\alpha}{}_{\beta}-
E^{\alpha}{}_{\beta}&=&\Delta^{\alpha}{}_{\beta}\,,
\label{second}
\end{eqnarray}
respectively, where $\Sigma_{\alpha}$ and $\Delta^{\alpha}{}_{\beta}$ are the
canonical energy--momentum and hypermomentum current three--forms
associated with matter. The left hand sides of
(\ref{first})--(\ref{second}) involve the gravitational gauge field
momenta two-forms $H_{\alpha}$ and $H^{\alpha}{}_{\beta}$.
We find them, together with
$M^{\alpha\beta}$, by partial differentiation of the Lagrangian
(\ref{QMA}):
\begin{eqnarray}
M^{\alpha\beta}&:=&-2{\partial V_{\rm MAG}\over \partial Q_{\alpha\beta}}
= -{2\over\kappa}\left\{{}^*\! \left(\sum_{I=1}^{4}b_{I}{}^{(I)}
Q^{\alpha\beta}\right)+{1\over2}\,
b_5\left[\vartheta^{(\alpha}\wedge{}^*(Q\wedge\vartheta^{\beta)})
-{1\over 4}\,g^{\alpha\beta}\,^*(3Q+\Lambda)\right]\right. \nonumber\\
&+& \left. c_{2}\,\vartheta^{(\alpha}\wedge{}^*{}\!^{(1)}T^{\beta)} +
c_{3}\,\vartheta^{(\alpha}\wedge{}^*{}\!^{(2)}T^{\beta)} + {1\over 4}
(c_{3}-c_{4})\,g^{\alpha\beta}{}^*\!\, T \right\}\, ,\label{M1}\\
H_{\alpha}&:=&-{\partial V_{\rm MAG}\over \partial T^{\alpha}} = -
{1\over\kappa}\, {}^*\!\left[\left(\sum_{I=1}^{3}a_{I}{}^{(I)}
T_{\alpha}\right) + \left(\sum_{I=2}^{4}c_{I}{}^{(I)}
Q_{\alpha\beta}\wedge\vartheta^{\beta}\right)\right],\label{Ha1}\\
H^{\alpha}{}_{\beta}&:=& - {\partial V_{\rm MAG}\over \partial
R_{\alpha}{}^{\beta}}= {a_0\over 2\kappa}\,\eta^{\alpha}{}_{\beta}
+ {\cal W}^{\alpha}{}_{\beta} + {\cal Z}^{\alpha}{}_{\beta},\label{Hab1}
\end{eqnarray}
where we introduced the abbreviations
\begin{equation}
{\cal W}_{\alpha\beta}:= {}^*\!
\left(\sum_{I=1}^{6}w_{I}{}^{(I)}W_{\alpha\beta} \right),\quad\quad
{\cal Z}_{\alpha\beta}:= {}^*\!
\left(\sum_{I=1}^{5}z_{I}{}^{(I)}Z_{\alpha\beta} \right).
\end{equation}
The three--forms $E_{\alpha}$ and $E^{\alpha}{}_{\beta}$
describe the canonical energy--mo\-men\-tum and hypermomentum
currents of the gauge fields themselves. They can be written as follows
\cite{PR}:
\begin{eqnarray}
E_{\alpha} & = & e_{\alpha}\rfloor V_{\rm MAG} + (e_{\alpha}\rfloor
T^{\beta})
\wedge H_{\beta} + (e_{\alpha}\rfloor R_{\beta}{}^{\gamma})\wedge
H^{\beta}{}_{\gamma} + {1\over 2}(e_{\alpha}\rfloor Q_{\beta\gamma})
M^{\beta\gamma}\\
E^{\alpha}{}_{\beta} & = & - \vartheta^{\alpha}\wedge H_{\beta} -
M^{\alpha}{}_{\beta}\, ,
\end{eqnarray}
where $e_{\alpha}\rfloor$ denotes the interior product with the
frame. We will restrict ourselves, from now on, only to the {\em  
electrovacuum} case
\begin{equation}
L=V_{\rm MAG}+V_{\rm Max}
\label{Ltot}\,,
\end{equation}
where $ V_{\rm Max}=-(1/2)F\wedge\hspace{-0.8em}  {\phantom{F}}^{\star}F$
is the Lagrangian of the Maxwell field and $F=dA$ is the electromagnetic field
strength. $\Delta^{\alpha}{}_{\beta}=0$ and
the only external current is the electromagnetic one
\begin{equation}
\Sigma_{\alpha}^{(max)}= 2 a_0 \,\left( e_\alpha\rfloor L_{{\rm Max}}
+(e_\alpha\rfloor F)\wedge H \right)
\label{max}\, .
\end{equation}
We concentrate ourselves on the simplest non--trivial case with shear.
We assume that nonmetricity and torsion are represented by a {\em
triplet of one--forms} \cite{Ob97,De96}, i.e., the Weyl covector (the dilation 
piece), the traceless covector piece of the nonmetricity
(a proper shear piece) and the torsion trace, so that
\begin{eqnarray}
Q_{\alpha\beta}&=&\,Q(u,v)\;g_{\alpha\beta}\, +
\frac{4}{9}\,\left(\vartheta_{(\alpha}e_{\beta )}\rfloor \Lambda(u,v)\,
-{1\over 4}g_{\alpha\beta}\Lambda(u,v)\,\right),
\label{nonmet}\\
T^\alpha&=&\frac{1}{3}\,\vartheta^\alpha\wedge T(u,v)\; , \qquad \hbox{with}
\qquad T:=e_{\alpha}\rfloor T^{\alpha}
\label{tor}\, .
\end{eqnarray}
Thus we are left with three non--trivial one--forms $Q(u,v)$,
$\Lambda(u,v)$, and $T(u,v)$.

We assume the following ansatz, the so--called {\em triplet ansatz}, for our
triplet of one forms (\ref{nonmet}) and (\ref{tor}):
\begin{equation}
Q=k_{0}\,\phi \,,\qquad \Lambda =k_{1}\,\phi \,,\qquad T=k_{2}\,\phi \,,
\label{tripp}
\end{equation}
where $k_{0}$, $k_{1}$, and $k_{2}$ are coupling constants. In
other words, we assume that the triplet of one--forms are proportional to
each other \cite{glmms98,hema99,Ob97,tw}.

The triplet ansatz (\ref{tripp}) reduces the MAG field equations
(\ref{first})--(\ref{second}) to an effective Einstein--Proca system  
\cite{Ob97,tw}:
\begin{eqnarray}
\frac{a_{0}}{2}\,\eta _{\alpha \beta \gamma }\wedge \tilde{R}^{\beta \gamma} 
&=&\kappa \,\Sigma _{\alpha }^{(\phi )},  \label{eq:Ein0} \\
d\,{}^{\ast }\!d\phi + m^{2}\,{}^{\ast }\!\phi &=&0,  \label{eq:Proca0}
\end{eqnarray}
with respect to the metric $g$, and the Proca 1--form $\phi $. Here the
tilde\,\,$\tilde{\null}$\,\, denotes the Riemannian part of the curvature.
$\Sigma _{\alpha }^{(\phi )}$ is the energy--momentum current of the Proca  
field $\phi$.
Moreover, by setting $m=0$ the system acquires a constraint among the coupling 
constants of the Lagrangian (\ref{lobo}), and the MAG system of field  
equations reduces
to the Einstein--Maxwell system, cf. Ref. \cite{hema99}.

Therefore, under the triplet ansatz (\ref{tripp}) and for the {\em  
electrovacuum} case
(\ref{Ltot}), the MAG field equations become
\begin{eqnarray}
\frac{a_{0}}{2}\,\eta _{\alpha \beta \gamma }\wedge \tilde{R}^{\beta \gamma} 
&=&\kappa \,\Sigma _{\alpha }^{(\phi )} + \Sigma_\alpha^{(max)},
\label{eq:Ein01} \\
d\,{}^{\ast }\!d\phi  &=&0,  \label{eq:Proca01}\\
d\,{}^{\ast }\!d A  &=&0,
\end{eqnarray}
with $\Sigma_\alpha^{(max)}$ given by (\ref{max}), and $A$ the electromagnetic 
one--form. Hence, the problem is reduced to find the metric, the electromagnetic
and the triplet one--forms.

\section{A class of cylindrically symmetric waves in MAG}

We start from the following coframe with coordinates $(t,x,y,\sigma)$:
\begin{eqnarray}
\vartheta^{\hat{0}}&=&\frac{1}{H}\sqrt{\frac{\widetilde{\Delta}}{\Upsilon}}  
\; dt\,,\\
\vartheta^{\hat{1}}&=&\frac{1}{H}\sqrt{\frac{\Upsilon}{\widetilde{\Delta}}} \;  
(dy+x^2d\sigma)\,,\\
\vartheta^{\hat{2}}&=&\frac{1}{H}\sqrt{\frac{\widetilde{\Delta}}{X}} \; dx \, , \\ 
\vartheta^{\hat{3}}&=&\frac{1}{H}\sqrt{\frac{X}{\widetilde{\Delta}}}
\; (dy-t^2d\sigma)
\label{coframe} \, .
\end{eqnarray}
Here we have the structure functions
$H=H(x,t)$, $X=X(x)$, $\Upsilon=\Upsilon(t)$, and $\widetilde\Delta =  
\widetilde\Delta(x,t)$. The coframe is assumed
to be orthonormal
\begin{equation}
g=o_{\alpha\beta}\,\vartheta^\alpha\otimes\vartheta^b\,.
\end{equation}
Then the metric explicitly reads
\begin{equation}
g=\frac{1}{H^2} \left \{- \frac{\widetilde{\Delta}}{\Upsilon}\, dt^2 +
\frac{\Upsilon}{\widetilde{\Delta}} \left( dy+ x^2 \, d \sigma\right)^2
+ \frac{\widetilde{\Delta}}{X}
\,dx^2 + \frac{X}{\widetilde{\Delta}} \left( dy - t^2 d \sigma
\right)^2 \right \}
\label{ortho} \, .
\end{equation}

When we substitute the local metric $o_{\alpha\beta}$, the coframe
(\ref{coframe}), the nonmetricity (\ref{nonmet}), and the torsion
(\ref{tor}) into the field equations (\ref{first}), (\ref{second}) of
the Lagrangian (\ref{lobo}), then, provided the constraints
\begin{eqnarray}
32 a_0^2 b_4 - 4a_0 a_2 b_4+64 a_0 b_3 b_4- 32 a_2 b_3 b_4 + 48 a_0 b_4 c_3
+ 24 b_4 c_3^2 + 24 b_3 c_4^2 & & \nonumber \\
+ 12 a_0 a_2 b_3 +48 a_0 b_3 c_4 - 9a_0 c_3^2 + 18 a_0 c_3 c_4 + 3a_0 c_4^2
+6a_0^2 a_2 + 24 a_0^2 c_4 & = & 0 \label{const1} \, , \\
b_5 &=& 0
\label{const} \, ,
\end{eqnarray}
on the coupling constants are fulfilled, we find a general exact
solution for the following functions:
\begin{equation}
\phi = \frac{H}{\sqrt{\widetilde{\Delta}}}
  \left(\frac{N_{{\rm e}}\,t}{\sqrt{\Upsilon}}\;\vartheta^{\hat{1}}+
    \frac{N_{{\rm g}}\,x}{\sqrt{X}}\;\vartheta^{\hat{3}}\right)
\label{sol1}\,,
\end{equation}
\begin{eqnarray}
H(x,t) &:=& 1 - \mu_0 x t \, , \nonumber\\
X(x) &:=& b - \left[g^2 + g_0^2\right] + 2 n x -
\epsilon_0 x^2 +2 m \mu_0 \, x^3 + \left( -\frac{\lambda}{3}- \mu_0^2 b -
\mu_0^2 \left[e^2 + e_0^2\right] \right)\, x^4 \, , \nonumber\\
\Upsilon(t) &:=&-(b+ \left[e^2 + e_0^2\right]e^2) + 2 m t - \epsilon_0 t^2
+ 2 n \mu_0 \, t^3
+ \left( \frac{\lambda}{3} + \mu_0^2 b -
\mu_0^2 \left[g^2 + g_0^2\right] \right)\, t^4 \, , \nonumber\\
\widetilde{\Delta}(x,t)&:=& x^2 + t^2 \, .
\label{solutions}
\end{eqnarray}
Here $N_{\rm e}$ and $N_{\rm g}$ are the gravito--electric and
gravito--magnetic nonmetricity--torsion charges of the source which
fulfill
\begin{equation}
\frac{z_4 k_0^2}{2a_0}\left( N_{\rm e}^2+N_{\rm g}^2 \right)= e^2+ g^2 \, .
\label{z4}
\end{equation}
The coefficients $k_{0}, k_{1}, k_{2}$ in (\ref{sol1}) are determined
by the dimensionless coupling constants of the Lagrangian (\ref{lobo}):
\begin{eqnarray}
k_0 &:=& \left({a_2\over 2}-a_0\right)(8b_3 + a_0) - 3(c_3 + a_0 )^2\,,
\label{k0}\\
k_1 &:=& -9\left[ a_0\left({a_2\over 2} - a_0\right) +
(c_3 + a_0 )(c_4 + a_0 )\right]\,,
\label{k1}\\
k_2 &:=& {3\over 2} \left[ 3a_0 (c_3 + a_0 ) + (
8b_3 + a_0)(c_4 + a_0 )\right]
\label{k2}\, .
\end{eqnarray}
Then, it is easy to put the constraint (\ref{const1}) into the following  
more compact
form
\begin{equation}
  b_4=\frac{a_0k+2c_4k_2}{8k_0}\,,\qquad\hbox{with}\qquad k:=
  3k_0-k_1+2k_2
\label{b4}\, .
\end{equation}
Therefore, the constants $\mu_0$, $g$, $e$, $b$, $\epsilon_0$, $m$, and $n$  
are free
parameters.

The electromagnetic potential $A$ associated to these solutions reads:
\begin{eqnarray}
{A}&=& \frac{1}{\widetilde {\Delta}}\left[( e_0\,x + g_0\,t)\, { dy}
+ (g_0\,x - e_0\,t)\, t \,x \,{ d \sigma} \right]\nonumber\\
&=& \frac{H}{\sqrt{\widetilde \Delta}}\left( \frac{g_0\, t}{\sqrt{\Upsilon}} \;
\vartheta^{\hat{1}} + \frac{e_0\, x}{\sqrt{X}}
\;\vartheta^{\hat{3}}\right)
\label{KK}\, ,
\end{eqnarray}
with $e_0$ and $g_0$ the electric and the magnetic charges, respectively.

Collecting our results, nonmetricity and torsion read:
\begin{equation}
Q^{\alpha\beta} = \left[k_0 \,o^{\alpha\beta} +\frac{4}{9}\, k_1
\,\left(\vartheta^{(\alpha}e^{\beta )}\rfloor-\frac{1}{4}\,
o^{\alpha\beta}\right)\right]\;
\frac{H}{\sqrt{\widetilde{\Delta}}} \left(\frac{N_{\rm e}\, t}{\sqrt{\Upsilon}}\;
\vartheta^{\hat{1}}+\frac{N_{\rm g}\,x}{\sqrt{X}}\;\vartheta^{\hat{3}}\right)
\label{nichtmetrizitaet}\, ,
\end{equation}
\begin{equation}
T^\alpha = \frac{k_2}{3}\;\vartheta^\alpha\wedge\;
\frac{H}{\sqrt{\widetilde{\Delta}}} \left(\frac{N_{\rm e}\,t}{\sqrt{\Upsilon}} \;
\vartheta^{\hat{1}}+\frac{N_{\rm g}\,x}{\sqrt{X}} \;\vartheta^{\hat{3}}\right)
\label{torsion} \, .
\end{equation}
The physical interpretation of the parameters of the solution, is clear:
The dilation (``Weyl") charges (related to
$^{(4)}Q_{\alpha\beta}$) are described by $k_0 N_e$ and $k_0 N_g$.
The shear charges  (related to $^{(3)}Q_{\alpha\beta}$) by $k_1 N_e$
and $(k_1 N_g)$, and, eventually, the  spin charges
(related to $^{(2)}T^\alpha$) by $(k_2 N_e)$ and $(k_2 N_g)$.

\section{Colliding waves in MAG}

At this point it is important to mention that not every cylindrically
symmetric spacetime, even in vacuum, satifies the colliding waves
requirements of Ernst \cite{ernst}, only certain classes of cylindrically
symmetric solutions can be thought of as solutions generated by collision
of waves.

The set of colliding waves solutions in metric--affine gravity is
described by the line element
\begin{equation}
g=2g(u,v) du dv + g_{ab}(u,v) dx^a dx^b \, , \qquad\, a,b=1,2\, ,
\end{equation}
where $x^1 = y$, $x^2=\sigma$ are ignorable coordinates.  The domain of the
coordinates chart consists of $(y,\sigma) \in {\bf R}^2$ and $(u,v)\in
{\bf R}^2$. As usual, it is the union of four continuous regions: $I:=
\{(u,v): 0\leq u < 1, 0\leq v < 1\}$ , $II:= \{(u,v): u\leq 0,
 0\leq v < 1\}$, $III:= \{(u,v): 0\leq u < 1, v < 0\}$,
$IV:=\{(u,v): u\leq 0, v\leq 0\}$.

We shall assume that the triplet one--form $\phi$ shares
the spacetime symmetries, i.e. it depends on the variables $u$ and
$v$ only. The metric and the triplet fields have to be continuous over the
whole domain.

In the region $IV$, a closed subregion of the Minkowski space, it
is required that $g(u,v)= g(0,0)$, $A = 0$, and $\phi=\, 0$.

In region $II$, the metric components  and the triplet one--form
depend only on $v$ i.e. $g=g(0,v)$, $A=A(0,v)$, and $\phi=\phi(0,v)$.
In region $III$ these fields are functions of the coordinate $u$,
i.e., $g_{\mu\nu}=g_{\mu\nu}(u,0)$, $A=A(u,0)$, and $\phi=\phi(u,0)$.

In region $I$, which is
occupied by the scattered null fields, the metric components, the electromagnetic, 
and the triplet one--forms are all functions of both $u$ and $v$  
coordinates.
In this way, the problem is reduced to know the metric (coframe), the
electromagnetic potential $A$, and the one--form $\phi$.

Moreover, until now our analysis has been concerned mostly with metrical
aspects leaving aside the task of having suitable field and matter
energy--momentum tensors. As it is well known, each participating wave in
the head--on collision is assumed to be plane fronted, i.e. it is
characterized by a covariantly constant null
eigenvector {\bf $k$}, i.e. $k_{\mu; \nu}=0$, of the Weyl conformal tensor of
Petrov type N. This condition implies a null energy--momentum
tensor for the electromagnetic field, i.e., a radiation field
or electrovacuum, into the field equations, c.f. Kramer et. al. \cite{Kram}.  
Therefore, the existence of colliding wave solutions with non--vanishing
cosmological $\lambda$ term is thus forbidden.

In this approach, one deals with real polynomials of fourth degree,
in order to express our solutions (\ref{sol1}) and (\ref{solutions}) in
terms of the advanced and retarded time coordinates ${\widetilde u}$,
${\widetilde v}$, one has to use the procedure outlined in Ref. \cite{bgmy98}.

When looking for advanced and retarded time variables one has to deal with
elliptic integrals,
i.e. we shall use the Legendre first kind integrals. The explicit
transformation of the elliptical integral to the Legendre first kind
integral is given by
\begin{equation}
\int \frac{dx}{\sqrt{G_4(x)}} = \mu \int \frac{d\phi}{\sqrt{1-k^2 \sin^2 \phi}}
=\mu F(\phi,k)\, ,
\end{equation}
where $G_4(x)$ is a fourth degree polynomial.
The real roots are denoted by $r_j$, with $j=1,\dots , 4$ and
$r_1 > r_2 > r_3 > r_4$, while the complex roots by $s_1 \pm i t_1$,
$s_2 \pm i t_2$, with $s_1 \ge s_2$, $t_1>t_2>0$, we introduce also the
following usefull notation
\begin{eqnarray}
r_{ik} &=& r_k - r_i, \, \quad (i,k=1,2,3,4)\, , \nonumber\\
\left(r,\beta,\gamma,\delta  \right) &=& \frac{r-\gamma}{r-\delta}\,
\frac{\beta-\delta}{\beta-\gamma}\, , \nonumber \\
\tan \theta_1 &=& \frac{r_1- s_1}{t_1}\, , \quad \tan \theta_2 =
\frac{r_2- s_1}{t_1}\, , \nonumber \\
\tan \theta_3 &=& \frac{t_1 + t_2}{s_1-s_2} \, , \quad
\tan \theta_4 = \frac{t_1 - t_2}{s_1-s_2}\, , \nonumber \\
\tan \left[(\theta_5/2)^2\right] &=& \frac{\cos \theta_3}{\cos \theta_4}\, ,
\nonumber \\
\nu &=& \tan \left[(\theta_2 - \theta_1)/2\right] \tan \left[(\theta_2 +
\theta_1)/2\right]\, .
\end{eqnarray}
The elliptic integral $F(\phi,k)$ is the Legendre integral of the first kind.
As it is well known, the standard form of writing this function is
\begin{equation}
z=\int^\phi \frac{d\phi}{\sqrt{1-k^2 \sin^2 \phi}}=F(\phi,k)\, , \quad
\phi = {\rm am}\, z
\label{titi}\, ,
\end{equation}
where ${\rm am}\, z$ denotes the function amplitude of $z$.
Replacing $\phi$ through $\omega$, according to
$\omega = \sin \phi= \sin {\rm am}z := {\rm sn}\, z$,
reduces Eq. (\ref{titi}) to
\begin{equation}
z= \int^\omega \frac{d\omega}{(1-\omega^2) \sqrt{1-k^2 \omega^2}}=
{\widetilde F}(\omega,k)\, ,
\end{equation}
where ${\rm sn}\, z$ belongs to the Jacobi family of elliptic functions
$({\rm sn}\, z, {\rm cn}\, z, {\rm dn}\, z)$, with well established analytical
properties. The values and main properties of these function can be found in 
tables, see for example \cite{korn}.

Let us return to our problem, the two dimensional line element can be written
in terms of the retarded and advanced time coordinates ${\widetilde u}$ and
${\widetilde v}$, respectively, as follows
\begin{equation}
\frac{dx^2}{X_4(x)} - \frac{dt^2}{\Upsilon_4(t)} =
d{\widetilde u}\otimes d{\widetilde v}
= 4 \frac{du}{U}\otimes \frac{dv}{V}\, ,
\end{equation}
with $U=\sqrt{1-u^2}$, and $V=\sqrt{1-v^2}$. Moreover,
\begin{equation}
\left. \begin{array}{l}
 d{\widetilde u}\cr
d{\widetilde v}
\end{array} \right \}  =\frac{dx}{\sqrt{X_4(x)}} \pm   
\frac{dt}{\sqrt{\Upsilon_4(t)}} = \mu_\phi
\frac{d\phi}{\sqrt{1-k_\phi^2 \sin^2 \phi}} \pm \mu_\theta
\frac{d\theta}{\sqrt{1-k_\theta^2 \sin^2 \theta}}
\label{beto1}\, ,
\end{equation}
where $u=\sin \frac{\widetilde u}{2}$, and $v=\sin \frac{\widetilde v}{2}$.
Straightforward integration of (\ref{beto1}) yields
\begin{equation}
\left. \begin{array}{l}
 {\widetilde u}\cr
{\widetilde v}
\end{array} \right \} = \mu_\phi F(\phi, k_\phi) \pm  \mu_\theta
F(\theta, k_\theta)
\label{beto2}\, .
\end{equation}
Hence, one has relations between the Legendre elliptic integrals and the
null coordinates $\widetilde u$ and $\widetilde v$, i.e.,
\begin{eqnarray}
F(\phi, k_\phi)&=& \frac{1}{2\mu_\phi}\left({\widetilde u}+{\widetilde v}
\right)=\arcsin \left(uV+vU\right)\, , \\
F(\theta, k_\theta) &=& \frac{1}{2\mu_\theta}\left(
{\widetilde u}-{\widetilde v}\right)=\arcsin \left(uV-vU\right)\, .
\end{eqnarray}
Therefore, the inversion formulas bring the following functional dependence
upon the generalized advanced and retarded time coordinates
\begin{eqnarray}
\phi&=& {\rm am} \left[\frac{1}{2\mu_\phi}\left({\widetilde u}+{\widetilde v}
\right)\right]\, , \quad \omega= \sin \phi = {\rm sn}
\left[\frac{1}{2\mu_\phi}\left({\widetilde u}+{\widetilde v}\right)\right]\, ,
\\
\theta&=& {\rm am} \left[\frac{1}{2\mu_\theta}\left({\widetilde u}-
{\widetilde v}\right)\right]\, , \quad \Omega= \sin \theta = {\rm sn}
\left[\frac{1}{2\mu_\theta}\left({\widetilde u}-{\widetilde v}\right)\right]
\, ,
\end{eqnarray}
which allow us to write the original $x$ and $t$ coordinates, initially
expressed through $\phi$ and $\theta$, in terms of the null coordinates
$\widetilde u$ and $\widetilde v$, i.e., $x=x({\widetilde u},{\widetilde v})$, 
and $t=t({\widetilde u},{\widetilde v})$.
One may encounter, in general, combinations of the different possible cases
depending on the character of the roots of the fourth degree polynomials,
namely, real or complex (For a detailed treatment of all possible cases,
see Ref.\cite{bgmy98}).

For instance, let us consider the case $X_4$ with four real roots and $\Upsilon_4$ 
also with four real roots, in which the coefficients of the higher degree are 
$1$. These conditions lead to constraints on the coefficients appearing
in the polynomials.
The roots of $X_4$ and $\Upsilon_4$ are all real and different, moreover, they
are denoted by $r_i$ and $\rho_i$ respectively. For the elliptic integral
depending on the $x$--coordinate, with $X_4(x)= (x-r_1)(x-r_2)(x-r_3)(x-r_4)$,
one has the relation
\begin{equation}
\int \frac{dx}{\sqrt{X_4(x)}} = \mu_\phi \int \frac{d\phi}{\sqrt{1-k_\phi^2
\sin^2 \phi}} = \mu_\phi F(\phi,k)\, ,
\end{equation}
where the explicit relation between $x$ and $\phi$ reads
\begin{eqnarray}
x&=& \frac{r_1 r_{42} - r_2 r_{41} \sin^2 \phi}{r_{42} - r_{41} \sin^2 \phi}\,
 , \nonumber \\
\sin^2 \phi &=& \frac{r_{42}}{r_{41}} \frac{x-r_1}{x-r_2}\, \quad \Rightarrow
\quad \phi = \arcsin \left(\pm \sqrt{\frac{r_{42}}{r_{41}} \frac{x-r_1}{x-r_2}}
\right)
\label{eureca}\, .
\end{eqnarray}
The parameter $k_\phi$ ($0<k_\phi^2<1$) is given by
\begin{equation}
k_\phi^2= (r_1, r_2, r_4, r_3) =  
\frac{r_1-r_4}{r_1-r_3}\frac{r_2-r_3}{r_2-r_4}\, ,
\quad {\rm and} \quad \mu_\phi = \frac{2}{\sqrt{r_{31}r_{42}}}\, .
\end{equation}
It is straightforward to find the analogous expression for the elliptic
integral depending on the $t$--coordinate.
The explicit expressions of the coordinates $x$ and $t$ in terms of  
${\widetilde u}$
and ${\widetilde v}$ are the following
\begin{eqnarray}
x({\widetilde u},{\widetilde v})&=& \frac{r_1 r_{42}
- r_2 r_{41}\left({\rm sn} \frac{1}{2\mu_\phi} ({\widetilde u} +{\widetilde v})
\right)^2}{r_{42} - r_{41}\left({\rm sn} \frac{1}{2\mu_\phi} ({\widetilde u} +
{\widetilde v})\right)^2} \label{x1}\, , \\
t({\widetilde u},{\widetilde v})&=& \frac{\rho_1 \rho_{42} - \rho_2
\rho_{41}\left({\rm sn} \frac{1}{2\mu_\theta} ({\widetilde u}
- {\widetilde v})\right)^2}{\rho_{42} - \rho_{41}
\left({\rm sn} \frac{1}{2\mu_\theta}({\widetilde u} - {\widetilde v})
\right)^2}
\label{y1}\, .
\end{eqnarray}

On the other hand, the relations between $\phi$ and $\theta$ with $x$ and $y$ 
read
\begin{equation}
\phi \pm \theta= \arcsin \sqrt{\frac{r_{42}}{r_{41}}\frac{x-r_1}{x-r_2}}  \pm
\arcsin \sqrt{\frac{\rho_{42}}{\rho_{41}} \frac{t-\rho_1}{t-\rho_2}}
\, ,
\end{equation}
where $\phi$ and $\theta$ stand correspondingly for the arguments of the
elliptical Legendre integrals related with the integration of $X_4$ and
$\Upsilon_4$. We define the auxiliary variables $u$ and $v$ by using the relation 
$\arcsin a \pm \arcsin b = \arcsin \left(a\sqrt{1-b^2} \pm b\sqrt{1-a^2}\right)$.
Thus,
\begin{equation}
2\arcsin u = \phi + \theta \, , \quad 2\arcsin v = \phi - \theta\, ,
\end{equation}
with
\begin{eqnarray}
\phi &=& \arcsin u + \arcsin v=\arcsin \left[u \sqrt{1-v^2}  + v \sqrt{1-u^2} 
\right]\, , \nonumber \\
&=& \arcsin \left[uV+vU\right]
\label{fi}\, ,
\end{eqnarray}
analagously
\begin{equation}
\theta = \arcsin \left[uV-vU\right]
\label{teta}\, ,
\end{equation}
where
\begin{equation}
V=\sqrt{1-v^2}\, , \quad  {\rm and}\quad  U=\sqrt{1-u^2}\, .
\end{equation}
Thus, subtituing (\ref{fi}) and (\ref{teta}) into (\ref{eureca}) it is
staightforward to find
\begin{eqnarray}
x(u,v)&=& \frac{r_1 r_{42} - r_2 r_{41}\left(uV+vU\right)^2}{r_{42} - r_{41}
\left(uV+vU\right)^2} \label{x}\, , \\
t(u,v)&=& \frac{\rho_1 \rho_{42} - \rho_2
\rho_{41}\left(uV-vU\right)^2}
{\rho_{42} - \rho_{41} \left(uV-vU\right)^2}
\label{y}\, .
\end{eqnarray}
Moreover, the advanced and retarded time coordinates ${\widetilde u}$ and
${\widetilde v}$ are related with the variables $u$ and $v$ through
\begin{eqnarray}
\left. \begin{array}{l}
d{\widetilde u}\cr
d{\widetilde v}
\end{array} \right \}&= &\left[\frac{\mu_\phi}{\sqrt{1-k_\phi^2 (uV+vU)^2}}
\pm  \frac{\mu_\theta}
{\sqrt{1-k_\theta^2 (uV-vU)^2}}\right] \frac{du}{U} \nonumber \\
&+&\left[\frac{\mu_\phi}{\sqrt{1-k_\phi^2 (uV+vU)^2}} \mp  \frac{\mu_\theta}
{\sqrt{1-k_\theta^2 (uV-vU)^2}}\right] \frac{dv}{V}
\label{null} \, .
\end{eqnarray}
According to our general procedure (\ref{null}) can be inverted, yielding
$u=u({\widetilde u},{\widetilde v})$ and $v=v({\widetilde u},{\widetilde v})$. 
The explicit representation requires the use of
tables for the Jacobi functions.

Our class of solutions (\ref{sol1}), (\ref{solutions}), and (\ref{KK})
defined in
region I, can be extended to the full
spacetime by introducing the Heaviside step function
\begin{equation}
\Theta ({\widetilde u})= \left\{\matrix{1\, , \quad {\widetilde u} \ge 0\cr
                  0\, , \quad {\widetilde u}<0}\right.\; ,
\end{equation}
with $\Theta^2({\widetilde u}) = \Theta({\widetilde u})$,
and replacing
${\widetilde u}\rightarrow \Theta({\widetilde u}){\widetilde u}$,
${\widetilde v} \rightarrow \Theta({\widetilde v}){\widetilde v}$. For the
case of real roots of the polynomials, treated
extensively in Ref. \cite{bgmy98}, we have explicitly for region II
\begin{eqnarray}
x({\widetilde u},{\widetilde v})&=& \frac{r_1 r_{42}
- r_2 r_{41}\left({\rm sn} \frac{1}{2\mu_\phi} {\widetilde v}
\right)^2}{r_{42} - r_{41}\left({\rm sn} \frac{1}{2\mu_\phi}
{\widetilde v}\right)^2} \label{x11}\, , \\
t({\widetilde u},{\widetilde v})&=& \frac{\rho_1 \rho_{42} - \rho_2
\rho_{41}\left({\rm sn} \frac{1}{2\mu_\theta}{\widetilde v}\right)^2}
{\rho_{42} - \rho_{41} \left({\rm sn} \frac{1}{2\mu_\theta}
{\widetilde v}\right)^2}
\label{y11}\, ,
\end{eqnarray}
and for region III
\begin{eqnarray}
x({\widetilde u},{\widetilde v})&=& \frac{r_1 r_{42}
- r_2 r_{41}\left({\rm sn} \frac{1}{2\mu_\phi} {\widetilde u}\right)^2}
{r_{42} - r_{41}\left({\rm sn} \frac{1}{2\mu_\phi}
{\widetilde u}\right)^2} \label{x111}\, , \\
t({\widetilde u},{\widetilde v})&=& \frac{\rho_1 \rho_{42} - \rho_2
\rho_{41}\left({\rm sn} \frac{1}{2\mu_\theta}{\widetilde u}\right)^2}
{\rho_{42} - \rho_{41}
\left({\rm sn} \frac{1}{2\mu_\theta}{\widetilde u}\right)^2}
\label{y111}\, ,
\end{eqnarray}
where $(1/2\mu_\phi) {\widetilde u}=(1/2\mu_\theta) {\widetilde u}= \arcsin u$,
and $(1/2\mu_\phi) {\widetilde v}=(1/2\mu_\theta) {\widetilde v}= \arcsin v$.

\section{Colliding waves for second degree polynomials}

In this section, we recover from our more general approach to colliding
waves with fourth degree polynomials the well known cases involving
only second degree polynomials.

In the limit of polynomials of second degree, i.e., $X_2$, $\Upsilon_2$, one has 
$k_\phi=k_\theta=0$, and $\mu_\phi, \mu_\theta \rightarrow 1$, hence one
recovers from (\ref{null}) the widely used standard retarded and advanced
time null coordinates
\begin{equation}
d{\widetilde u}=2\frac{du}{U}\, \quad d{\widetilde v}=2\frac{dv}{V}
\label{null1}\, ,
\end{equation}
see below the example.

Let us restrict to the case $\mu=0$, $\lambda=0$, and $\epsilon=1$, with this 
choice the fourth degree polynomials $X_4$ and $\Upsilon_4$ reduce to second  
degree
ones $X_2$ and $\Upsilon_2$, respectively. Therefore, these polynomials now read 
\cite{alberto}:
\begin{eqnarray}
\Upsilon_2&=& - \left(t-m \right)^2+m^2 - b - \left[e^2 + e_0^2\right] = a^2
- (t-m)^2\, , \nonumber \\
&=& a^2 \left\{ 1- (uV-vU)^2 \right\} =a^2 \left[UV+ uv\right]^2
\end{eqnarray}
with $t-m =a (uV- vU)$, $a^2 :=  m^2 - b - \left[e^2 + e_0^2\right]$ and
\begin{eqnarray}
X_2&=&b - \left[g^2 + g_0^2\right] +2nx-x^2= h^2 - (x-n)^2\nonumber \, , \\
&=& h^2 \left\{ 1- (uV+vU)^2 \right\}=h^2 \left\{ UV-uv \right\}^2 \, .
\end{eqnarray}
with $x-n=h (uV+ vU)$, and $h^2:=n^2+ b - \left[g^2 + g_0^2\right]$.
In order to write the two dimensional line element we note that
\begin{equation}
dx = h(UV-uv) \left\{ \frac{du}{U} + \frac{dv}{V} \right\}\, , \quad
dt = a (UV+vu) \left\{ \frac{du}{U} - \frac{dv}{V} \right\}\, ,
\end{equation}
therefore
\begin{equation}
\frac{dx^2}{X} - \frac{dt^2}{\Upsilon}=4 \frac{du}{U} \frac{dv}{V} \, ,
\end{equation}
and
\begin{equation}
{\widetilde \Delta}=x^2 + t^2=\left[n+h(uV+vU)\right]^2 + \left[m+a(uV-vU)
\right]^2
\end{equation}
In this way, one arrives at a class of colliding wave solutions with second
degree polynomials, namely
\begin{eqnarray}
g &=& 4{\widetilde \Delta} \frac{du}{U} \frac{dv}{V} +
\frac{b^2(UV-vu)^2}{\widetilde \Delta} \left\{ dy -
\left[ m+a (uV-vU) \right]^2 d \sigma \right\}^2 \nonumber \\
&+& \frac{a^2 (UV+uv)^2}{\widetilde \Delta} \left\{ dy +\left[ n+h
(uV+vU)\right]^2 d\sigma \right\}^2
\label{plebcar}\, .
\end{eqnarray}
The corresponding electromagnetic potential reads
\begin{equation}
A=\frac{H}{\sqrt{\widetilde{\Delta}}}
\left(\frac{\left[m+a(uV-vU)\right]}{\sqrt{\Upsilon }}\;
\vartheta^{\hat{1}}+\frac{\left[n+h(uV+vU)\right]}{\sqrt{X}}\;
\vartheta^{\hat{3}}\right)\, .
\end{equation}
Then nonmetricity and torsion read
\begin{eqnarray}
Q^{\alpha\beta}&=&\left[k_0 \, o^{\alpha\beta}+\frac{4}{9}\,
k_1 \,\left(\vartheta^{(\alpha}e^{\beta )}\rfloor-\frac{1}{4}\,
o^{\alpha\beta}\right)\right] \frac{H}{\sqrt{\widetilde{\Delta}}}
\left(\frac{N_{\rm e}\, \left[m+a(uV-vU)\right]}{\sqrt{\Upsilon }}\;
\vartheta^{\hat{1}} \right.\nonumber \\
&+&\left. \frac{N_{\rm g}\, \left[n+h(uV+vU)\right]}{\sqrt{X}}\;
\vartheta^{\hat{3}}\right)
\label{nichtmet} \, ,
\end{eqnarray}
\begin{equation}
T^\alpha= \frac{k_2}{3} \vartheta^\alpha\wedge
\frac{H}{\sqrt{\widetilde{\Delta}}}
\left(\frac{N_{\rm e}\, \left[m+a(uV-vU)\right]}{\sqrt{\Upsilon }}\;
\vartheta^{\hat{1}}+\frac{N_{\rm g}\, \left[n+h(uV+vU)\right]}{\sqrt{X}}\;
\vartheta^{\hat{3}}\right)
\label{torsion1}\, ,
\end{equation}
respectively. The relation (\ref{z4}) for $z_4$ still stands.

As it has been pointed out, this solution describes the
scattering of two noncollinear polarized gravitation plane waves.  At
the leading edge of each colliding type--N gravitational wave, the
curvature tensor exhibits delta and jump discontinuities.
The former is interpreted as a gravitational impulsive
wave, whereas the latter is attributed to a gravitational shock wave.

As stated above, this class of solutions, defined again in region I, can also 
be extended to the full spacetime by introducing the Heaviside step function 
$\Theta$.

The nonmetricity and torsion present delta singularities and
jump discontinuities. However, the Bianchi identities hold in a distributional
sense, see \cite{taub}.
In particular, also $D T^\alpha = R_\beta^{\phantom{\alpha}\alpha}\wedge
\vartheta^\beta$ holds.
There are no problems on the right--hand side because the delta type
singularities of the curvature are multiplied by the smooth distributions
$\sqrt{1 - \Theta(u) u^2}$ and $\sqrt{1 - \Theta(v) v^2}$, respectively.

This example shows very clear the generality of our new approach, which
allows the interpretation of the resulting waves as curvature, nonmetricity
and torsion shock waves.

These solutions were checked with Reduce \cite{REDUCE} with its Excalc
package \cite{EXCALC} for treating exterior differential forms\cite{Stauffer}
and the Reduce--based GRG computer algebra system \cite{GRG,cpc}.

\section{Outlook}

Examples in which spacetime might become non--Riemannian above
Planck energies occur in string theory or in the very early universe in the
inflationary model. The simplest such geometry is metric--affine geometry, in
which nonmetricity appears as a field strength, side by side with curvature
and torsion \cite{nehe}.
Nowadays, there exist a revival of interest in metric affine gravity (MAG)
theories. It has been demostrated that they contain the axi--dilatonic sector 
of low energy string theory \cite{dt95} as special case. Moreover, the
gravitational interactions involving the axion and dilaton may be derived from
a geometrical action principle involving the curvature scalar with a
non--Riemannian connection. In other words, the axi--dilatonic sector of the 
low energy string theory can be expressed in terms of a geometry with torsion 
and nonmetricity \cite{dot95}. This formulation emphasizes the geometrical
nature of the axion and dilaton fields and raises questions about the most
appropiate geometry for the discussion of physical phenomena involving these 
fields.

It is important to mention that on the one hand the axion--dilaton theory
comes, as mentioned above, from the low energy limit of string models.
On the other hand such kind of models represents one sector of
the MAG models. Since these two models have one important sector in common,
we should consider the MAG models in a new perspective as an effective low
energy theory towards quantum gravity.


\section*{Acknowledgments}
We thank Friedrich W. Hehl, Yuri N. Obukhov, and Eckehard W. Mielke for
useful discussions and literature hints. This research was partially
supported by CONACyT Grants 28339E, 32138E, by the joint
German--Mexican project DLR--Conacyt MXI 010/98 OTH --- E130--1148,
and by FOMES Grant: P/FOMES 98--35--15.

\end{document}